\documentclass[showpacs,aps,prb,twocolumn]{revtex4}
\usepackage{graphicx}
\usepackage{amsmath}
\usepackage{amssymb}

\begin{document}

\bibliographystyle{apsrev}
\title{Anisotropic flux creep in Bi2212:Pb single crystal in crossed magnetic fields}

\author{L.~S.~Uspenskaya and A.~B.~Kulakov}
\affiliation{Institute of Solid State Physics, Russian Academy of
Sciences, Chernogolovka, Moscow Distr., 142432, Russia, e-mail:uspenska@issp.ac.ru}

\author{A.~L.~Rakhmanov}
\affiliation{Institute for Theoretical and Applied
Electrodynamics, Russian Academy of Sciences, Izhorskaya Str.
13/19, Moscow, 125412 Russia}
\date{\today}

\begin{abstract}
An experimental study of magnetic flux penetration under crossed
magnetic fields in Bi2212:Pb single crystals is performed by the
magneto-optic technique. The anisotropy of the flux creep rate
induced by the in-plane magnetic field is observed at $T<54\pm
2$~K. This observation confirms the existence of the
three-dimensional flux line structure in Bi2212:Pb at low
temperatures. An asymmetry of the flux relaxation with respect to
the direction of the in-plane field is found. This effect can be
attributed to the influence of the laminar structure on the
pinning in Bi2212:Pb single crystals.
\end{abstract}

\pacs{74.72.Hs, 74.60.Ge, 74.60.Jg, 74.25.Ha}
\smallskip

\maketitle

\section{Introduction} \label{In}

The magneto-optic (MO) study of the dynamics of the magnetic flux
in type-II superconductors in crossed magnetic fields is a
convenient tool for the investigation of the vortex lines
properties. In particular, the MO studies in crossed magnetic
fields are employed to clarify the presence or absence of
three-dimensional (3D) correlations in FLL of
superconductors.~\cite{In,Zeld,InvM,Vl-Vl,Grig,Tokun1,Tokun2,Mats}
In our previous papers a strong magnetic field induced anisotropy
was revealed in the single crystals of
(Bi$_{0.65}$Pb$_{0.35}$)$_{2.2}$Sr$_2$CaCu$_{2}$O$_{8+\delta}$
(Bi2212:Pb) within the temperature range $T < T_m = 54\pm
2$~K.~\cite{Usp1,Usp2} In these experiments, a plate like specimen
of a single crystal is placed in a DC magnetic field directed in
the sample plane, $\mathbf{H}_{ab}$, and then a field,
$\mathbf{H}_{z}$, perpendicular to the plane  is applied. In such
geometry the MO technique is used to study the penetration of the
magnetic flux induced by the field $\mathbf{H}_{z}$. The
experiments reveal two strikingly different types of flux
behavior. The transverse flux moves into the Bi2212:Pb single
crystals preferably along the direction of the in-plane magnetic
field $\mathbf{H}_{ab}$ if $T < T_m$. This type of the behavior is
analogous to that observed in YBCO single crystals~\cite{In} and
gives an evidence for the existence of the strong superconducting
correlations between CuO planes in Bi2212:Pb at $T < T_m$. Quite
the contrary, the transverse magnetic flux penetrates independent
of the orientation of the in-plane magnetic field  at $T > T_m$.
Such a behavior is observed in undoped Bi2212 single crystals and
indicates that flux lines in this system can be treated as 2D
pancakes.~\cite{In}

\begin{figure}
\includegraphics [width=80mm]{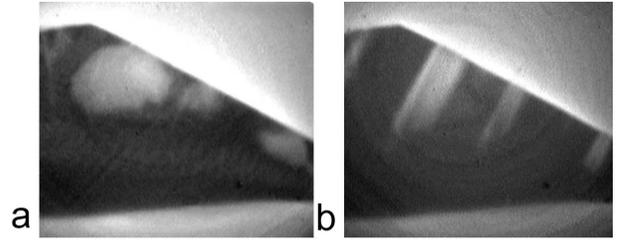}

\caption{\label{fig1} 'Bubble' and 'stripe' picture of $H_z$ field
penetration; $T=36$~K, $H_z = 77$~Oe, $H_{ab}=0$~Oe (a) and
$H_{ab}=650$~Oe (b).}
\end{figure}

The in-plane field induced flux penetration anisotropy is observed
in the 'ideal' crystals with uniform magnetic flux entering
through the sample edges. However, this effect is the most vivid
and demonstrative in the case of the crystal with strong defects
(or weak points) near its edges.~\cite{Usp1,Usp2} In the studied
Bi2212:Pb single crystals these defects were observed in the
points where the twin boundaries cross the sample edges. The
picture of the flux penetration near such defects is entirely
reproducible from the test to test. The magnetic flux enters the
sample near weak points in the form of some 'bubbles' or 'stripes'
which remain attached to the weak point during the time of
observation, Fig.~\ref{fig1}. The transverse magnetic flux has a
shape of the bubble if the in-plane magnetic field $H_{ab}=0$. The
bubble stretches along $\mathbf{H}_{ab}$ and shapes of a stripe if
$H_{ab}>0$.

The anisotropy of transverse flux penetration is characterized by
the 'geometric' factor (the evolution of the penetrated region
from the bubble to stripe), by the screening current anisotropy
(the current density along $\mathbf{H}_{ab}$ is much larger than
the current density across this direction), and by the flux creep
rate anisotropy as well. In the present paper we analyze a
peculiarity of the flux creep anisotropy in Bi2212:Pb single
crystals in crossed magnetic fields by means of the MO imaging.

\section{Experimental} \label{Ex}

The studied samples were Bi2212 single crystals doped by 35$\%$ of
lead and with the optimal oxygen content. The same samples we used
for MO imaging in our previous experiments.~\cite{Usp1,Usp2} More
detailed samples description is presented in
Refs.~\onlinecite{Kul,Kul1}. The visualization of the transverse
with respect to the sample surface magnetic flux component was
performed by the conventional real time MO technique.~\cite{MO}
The graphs of the magnetic flux distribution were obtained by
means of the MO images calibration procedure described in
Ref.~\onlinecite{Usp2}.

\begin{figure}
\includegraphics [width=40mm]{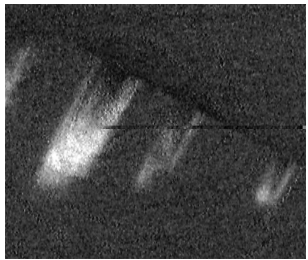}

\caption{\label{fig2} Relaxation of penetrated magnetic flux in
the case shown in Fig.~\ref{fig1}(b). The image is obtained by
substraction of two images taken at $t_1=0.2$~s and at
$t_2=90$~s.}
\end{figure}

The typical evolution of the magnetic flux distribution with time
$t$ is illustrated by Fig.~\ref{fig2}. The sample at temperature
higher than critical temperature $T_c$ was placed in the in-plane
magnetic $H_{ab}=650$~Oe directed perpendicular to one of the
crystal edge. Than the sample was cooled to $T=36$~K and the
transverse magnetic field $H_z = 77$~Oe was turned on. The picture
in Fig.~\ref{fig2} is obtained by substraction of two images. The
first image was recorded at $t_1=0.2$~s after turning on the field
$H_z$ and the second one at $t_2=90$~s. So, the bright region
corresponds to the volume into which the magnetic flux penetrates
during a time $\Delta t=t_2-t_1$. Naturally, some decrease of the
magnetic field occur near the sample edge (see dark region of the
image). It is seen from Fig.~\ref{fig2} that the magnetic flux
drifts along the vector $\mathbf{H}_{ab}$ significantly faster
than across it.

\begin{figure}
\includegraphics [width=60mm]{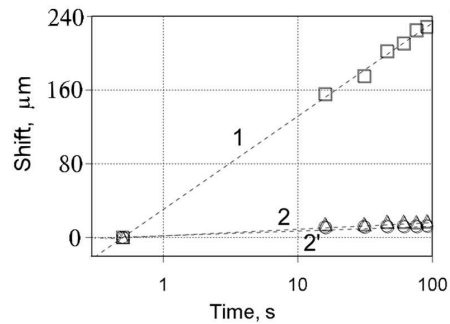}

\caption{\label{fig3} Evolution of flux penetration depth in
direction along (1) and across the field $\mathbf{H}_{ab}$ to the
left (2) and to the right ($2^{\prime}$); $T=30$~K, $H_z =
116$~Oe, and $H_{ab}=650$~Oe.}
\end{figure}

The time variation of the flux penetration depth is illustrated by
curves 1 and 2,$2^{\prime}$ in Fig.~\ref{fig3}. The first point
corresponds to the time moment about 0.1~s after turning on the
field $H_z$. The curve~1 shows the change with time of the
penetration depth $l_{\parallel}$ along the in-plane magnetic
field direction and curve~2 presents the change of penetration
depth $l_{\perp}$ across this direction. The creep anisotropy
$k_l=l_{\parallel}/l_{\perp}$ increases with the value of
$H_{ab}$. At fixed magnetic fields, this coefficient increases
with temperature if $T<T_m=54\pm 2$~K. The magnetic flux
penetration anisotropy at $t\approx 0$ also increases within the
same temperature range and the magnetic field induced anisotropy
disappears if $T>T_m$.~\cite{Usp1,Usp2}

\begin{figure}
\includegraphics [width=60mm]{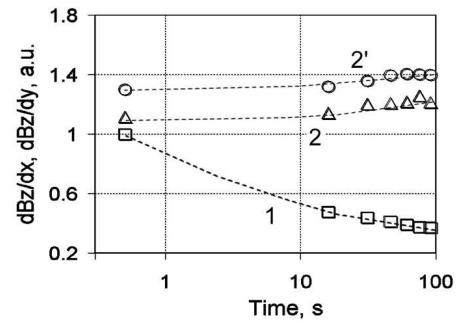}

\caption{\label{fig4} Relaxation of screening currents across (1)
and along the field $\mathbf{H}_{ab}$ at the left (2) and at the
right ($2^{\prime}$) side of the stripe. The values of temperature
and magnetic fields are the same as in Fig.~\ref{fig3}.}
\end{figure}

The magnetic flux entering in crossed fields is characterized also
by the anisotropy of the screening currents. The current
anisotropy increases with the increase of the in-plane magnetic
field and temperature if $T<T_m$.~\cite{Usp1,Usp2} For our
crystals, the ratio of the current density along and across the
direction of $\mathbf{H}_{ab}$, $k_J$ achieves the value up to
10--15 at $H_{ab}=1800$~Oe. The magnetic flux creep results not
only in the dipper flux penetration into the sample with time but
also in the relaxation of the screening currents. The current
density decay rate is different for different current components.
The values $\partial B_z/\partial y \propto j_{\perp}$ (curve~1)
and $\partial B_z/\partial x \propto j_{\parallel}$
(curves~2,$2^{\prime}$) versus time are shown in Fig.~\ref{fig4},
where $x$ is a coordinate axis across the in-plane field direction
and $y$ is a longitudinal one. The magnetic field derivatives were
taken at the 'stripe' peripheral where these values are almost
constant (see the end of this section). The relaxation behavior of
the currents along and across the direction of the vector $H_{ab}$
is quite different. First, the relaxation rate of the longitudinal
current is smaller than of the transverse one, as it follows from
the figure. Moreover, the longitudinal current increases slightly
with time which is rather unusual, while a common for the
screening current decay is observed for the current component
across the direction of the in-plane magnetic field.

Note also an interesting observation. The relaxation rate to the
right and to the left with the respect of the stripe direction is
different. This fact is indicated in Fig.~\ref{fig4} by curves 2
and $2^{\prime}$, which corresponds to the screening currents
relaxation to the left and to the right direction with respect to
the vector $\mathbf{H_{ab}}$ respectively. Their absolute values
and relaxation rates are slightly different. As it will be seen,
the observed difference is due to the inclination of the in-plane
field from the normal to the crystal edge. The inclination angle
$\alpha$ is about $6^{\circ}$ for the experimental data presented
in Fig.~\ref{fig4}. The difference in the creep rates increases
with $\alpha$ at any rate in the range $0<\alpha<45^{\circ}$. At
higher deviations of the in-plane field from the normal to the
edge the flux stripes corresponding different weak points begin to
overlap and the MO observation of the effect becomes difficult.

\begin{figure}
\includegraphics [width=80mm]{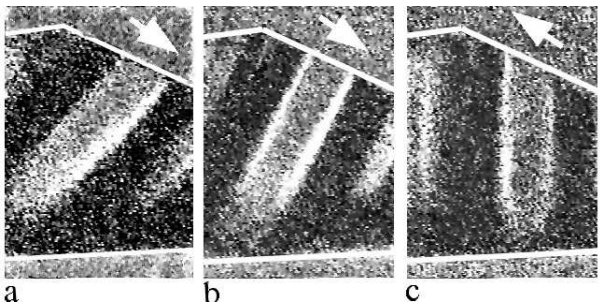}

\caption{\label{fig5} The evolution of flux pattern in 30~s
($T=30$~K, $H_z=116$~Oe, and $H_{ab}=1800$~Oe) for different
angles between $\mathbf{H}_{ab}$ and sample edge; $\alpha =
17^\circ$ (a), 2$^\circ$ (b), and -23$^\circ$ (c). The pictures
are obtained by substraction of images taken at $t_1=1$~s and
$t_2=30$~s. The sample edge are pencilled by white line. Arrows
show the direction of the flux preferential drift.}
\end{figure}

The asymmetry of the relaxation rate is clearly seen even at
relatively small deviation of the field $\mathbf{H}_{ab}$ from the
normal to the sample edge. Fig.~\ref{fig5} illustrates the
relaxation of the magnetic flux for the inclination angles
$\alpha=17^\circ$, 2$^\circ$, and -23$^\circ$ ($T=30$~K,
$H_z=116$~Oe, and $H_{ab}=1800$~Oe). The pictures were obtained by
the subtraction of two MO images taken at $t_1=1$~s and $t_2=30$~s
after turning on the field $H_z$. It should be emphasized that at
the first moment after turning on the transverse magnetic field,
the penetration lengths and screening currents at the right and
left parts of the flux front were equal (with the experimental
accuracy, of course). The asymmetry arises in the course of the
relaxation process.

The asymmetry changes its sign if one of the fields $H_z$ or
$H_{ab}$ changes the sign. Naturally, the change of the sign of
the both of the fields does not affect the magnetic relaxation
asymmetry. The direction of the preferable flux relaxation changes
also if the sign of the angle $\alpha$ between the in-plane field
and the normal to the specimen edge changes (see Fig.~\ref{fig5}).

\begin{figure}
\includegraphics [width=50mm]{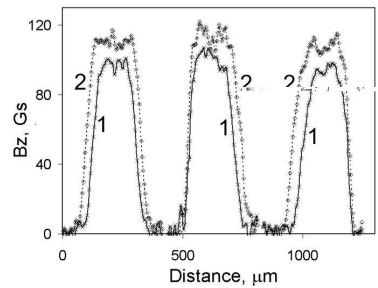}

\caption{\label{fig6} The evolution of induction profiles across
$\mathbf{H}_{ab}$ in 30~s for $\alpha = 2$, 17, and -23$^\circ$
($T=30$~K, $H_z=116$~Oe, and $H_{ab}=1800$~Oe). Curves 1
correspond to initial flux distribution and curves 2 correspond to
final one. The profiles are taken near the middle length of the
stripe.}
\end{figure}

Fig.~\ref{fig6} shows the magnetic field profiles across the
direction of the in-plane field taken at different moments of
times for three different angles $\alpha$ at $T=30$~K and
$H_{ab}=1800$~Oe. It follows from the figure, the flux penetration
asymmetry increases rapidly with time and grows (but not too fast)
with the inclination angle $\alpha$.

\section{Discussion} \label{Disc}

Our previous MO studies of Bi2212:Pb single crystals in crossed
magnetic fields revealed that the transition occurs in the
magnetic flux behavior at $T=T_m=54\pm2$~K.~\cite{Usp1,Usp2} The
transverse magnetic flux at $T<T_m$ behaves like in YBCO spreading
preferably along the in-plane magnetic field. At $T>T_m$ the
transverse flux penetrates independent of the in-plane magnetic
field as in Bi2212 system. The obtained results can be understood
within the concept of the flux line melting giving rise to the
transition of 3D correlated stacks of pancakes at $T<T_m$ into a
disordered phase of 2D ones at $T>T_m$.

The results obtained in the present study confirms the existence
of strong 3D correlations in the flux line structure in Bi2212:Pb
at $T<T_m$. Really, the creep assistant penetration rate along the
applied in-plane magnetic field is significantly higher than that
across this direction, that is, the activation barrier for the
flux line penetration transverse to $H_{ab}$ is higher than the
barrier along it. The fact that the relaxation rate of the
screening currents along the in-plane field is lower than the
current relaxation in the perpendicular direction supports the
above conclusion. In our experiments we cool the sample in the
in-plane field (field cooled regime). Under such a condition,
the in-plane structure of the flux lines arises in the sample. The
in-plane vortices should evidently locate preferably between CuO
planes.~\cite{Blat} The applied transverse magnetic field
$\mathbf{H}_z$, induces the entering of the flux lines directed
transverse with respect to the sample plane. The penetrating flux
lines should intersect the in-plane vortices when moving in the
direction perpendicular to $H_{ab}$. In the case of highly
anisotropic (layered) superconductors the transverse to the layers
magnetic flux has a form of 2D pancake-like
vortices.~\cite{Blat,Brandt,Fein} As a result, in such extremely
anisotropic superconductors as Bi2212, the transverse magnetic
flux enters the sample volume independently of the density and
direction of the in-plane vortices.~\cite{In} On the contrary, in
the less anisotropic systems such as in YBCO, the transverse
magnetic flux penetrates preferably along the direction of the
in-plane vortices since due to the strong superconducting coupling
between CuO planes there exists an additional energy barrier for
the intersection of the in-plane and transverse vortices. Thus, we
conclude that in Bi2212:Pb single crystals the pancakes in
different CuO planes are strongly correlated at $T<T_m$.

Some unusual effect is observed in transverse magnetic flux
relaxation. The magnetic field gradient across the in-plane field
direction increases with time, then, the corresponding component
of the screening current increases also. This effect could be
explained as follows. The entering of the transverse magnetic flux
near a weak point should be evidently accompanied by some
redistribution of the in-plane magnetic induction. The in-plane
field prevents the transverse vortex penetration. Thus, due to the
thermoactivated flux lines flow (TAFF), a part of the in-plane
vortices should be expelled from the region near the weak point to
the periphery of the region penetrated by the transverse flux. As
a result in this region some thickening of the in-plane vortices
arises with time. The current screening the flux penetration
across the direction of $\mathbf{H}_{ab}$ growths with the value
of the in-plane magnetic induction.~\cite{Usp1,Usp2} The described
effect could be a reason for the found increase with time of the
value $\partial B_z/\partial x \propto j_{\parallel}$ (see
Fig.~\ref{fig4} (curve~2)).

A rather interesting feature is the asymmetry of the flux creep
rate with respect to the in-plane field direction described at the
end of the previous section. This asymmetry can be attributed to
the peculiarities of the defect structure of the studied samples.
There are two types of characteristic planar defects in Bi2212:Pb
system. The first one is system of twins. These defects are seen
in polarized light and serves the weak points for the flux
penetration in the studied single crystals. The second kind of
planar defects in Bi2212:Pb is a so-called laminar
structure.\cite{Pb} These defects revealed by means of X-ray
diffraction are formed by modulation of the Pb concentration. The
laminar structure contributes to the total pinning force and the
screening current density along the laminae is higher than the
current density across them. In our samples the laminae are
parallel to two of the sample edges and the current anisotropy
along and across the laminae is about 1.5--2, which is
significantly lower than the in-plane magnetic field induced
anisotropy.~\cite{Usp1,Usp2} After the transverse field turning
on, the flux penetrates the sample preferably along the in-plane
magnetic field and later the smaller anisotropy due to lamina
structure proves itself.

The appearance of the asymmetry could be understood as follows.
The interaction of the transverse vortices with the currents
flowing in the superconductor gives rise to the existence of the
forces which drives the magnetic flux into the sample bulk. The
directions along the in-plane magnetic field and along the laminae
are more favorable for the flux line motion. In the case of the
in-plane field directed normal to the sample edge and,
consequently, perpendicular to the laminar structure, the motion
of the transverse flux to the left and to the right with respect
to $H_{ab}$ are equal. If there exists some angle $\alpha$ between
the in-plane field and the normal to the laminar structure, the
left and right directions become unequal since the component of
the driving force along the 'easy' direction (parallel to the
laminae) is different. This could give rise to the observed
asymmetry. For illustration, we calculate below the value of the
flux creep rate asymmetry using a simple model of TAFF.

Let us introduce a coordinate system $x,y,z$ with $z$ axis
directed perpendicular the sample plane, $y$ axis along the
in-plane field, and $x$ axis across it. The magnetic field has two
components $B_z$ and $B_y$ and the components of the current in
$ab$ plane are
\begin{equation}\label{j}
j_x = \frac{c}{4\pi}\left(\frac{\partial B_z}{\partial y}+
\frac{\partial B_y}{\partial z}\right), \,\,\ j_y =
-\frac{c}{4\pi}\frac{\partial B_z}{\partial x}\,.
\end{equation}
The components of the Lorentz's force acting on the transverse
flux line can be written as follows
\begin{equation}\label{fpar}
f_{\parallel}=\frac{\phi_0 n_z}{4\pi}\left[\frac{\partial
B_z}{\partial x}\sin{\alpha}-\left(\frac{\partial B_z}{\partial
y}+ \frac{\partial B_y}{\partial z}\right)\cos{\alpha}\right],
\end{equation}
\begin{equation}\label{ftr}
f_{\perp}=-\frac{\phi_0 n_z}{4\pi}\left[\frac{\partial
B_z}{\partial x}\cos{\alpha}+\left(\frac{\partial B_z}{\partial
y}+ \frac{\partial B_y}{\partial z}\right)\sin{\alpha}\right],
\end{equation}
where $\phi_0$ is the flux quantum, $n_z=1$ if the transverse
field $B_z$ is directed in the positive direction and $n_z=-1$ if
$B_z$ is directed in the negative one, the value $f_{\parallel}$
corresponds to the force component along the laminar structure,
and $f_{\perp}$ is the force component  across the laminae.
Following a standard approach~\cite {Blat}, we express the line
velocity as a sum of TAFF probabilities for the flux line to move
through a distance $l$ during a time interval $\tau$ down and
against the Lorentz force. As a result, we get for the components
of the flux line velocity along and across the laminar structure
\begin{equation}\label{prob}
v_i=2v_0\exp\left(-\frac{V_i}{kT} \right)\sinh{\left(\frac{f_i
lL_z}{kT}\right)},
\end{equation}
where $v_0=l/\tau$, $i=\parallel$ or $\perp$, $L_z$ is the flux
line length along $z$ axis, and $V_i$ are the effective pinning
barriers for the flux line motion along and across the laminar
structure. The component of the TAFF velocity $v_x$ transverse to
the vector of the in-plane field is defined now by evident formula
$v_x=v_{\parallel}\sin{\alpha}-v_{\perp}\cos{\alpha}$. At this
point we could start with the analysis of the flux penetration
asymmetry. However, for simplicity we linearized Eq.~(\ref{prob})
with respect to $f_i$ assuming $f_i lL_z/kT\ll 1$ and find
\[
v_x=\gamma n_z\left[\frac{\partial B_z}{\partial
x}\left(\sin^2{\alpha}+\beta\cos^2{\alpha}\right)- \right.
\]
\begin{equation}\label{transverse}
\left. \frac{1-\beta}{2}\left(\frac{\partial B_z }{\partial y
}+\frac{\partial B_y }{\partial z }\right)\sin{2\alpha}\right],
\end{equation}
where
\[
\gamma=\frac{v_0\phi_0 lL_z}{2\pi
kT}\exp{\left(-\frac{V_{\parallel}}{kT}\right)},
\]
\begin{equation}\label{not}
\beta=\exp{\left(\frac{V_{\parallel}-V_{\perp}}{kT}\right)}\leq 1.
\end{equation}

The sign of the derivative $\partial B_z/\partial x$ is evidently
different for the flux lines at the right and at the left side of
the magnetic flux penetration front. In the case of the isotropic
superconductor ($\beta=1$) or if the in-plane magnetic field is
directed perpendicular to the sample edge (that is,
$\alpha=\pi/2$), these lines moves in opposite directions with the
same absolute values of the velocity components $|v_x|$. If there
exists in-plane anisotropy and the in-plane magnetic field
deviates from the normal to the edge, then, the absolute values of
the transverse velocity components are different and from
Eq.~(\ref{transverse}) we find
\[
\Delta v_x=|v_x^{right}|-|v_x^{left}|=
\]
\begin{equation}\label{asym}
\gamma (1-\beta)n_z\left(\frac{\partial B_z}{\partial
y}+\frac{\partial B_y}{\partial z}\right)\sin{2\alpha}.
\end{equation}
This asymmetry disappears at $\alpha = \pi/2$ as it is observed in
the experiment. The obtained result is based on the simplest
possible model of TAFF and could not be used for a quantitative
analysis. Nevertheless, it describes the main features of the
effect since the observed flux motion asymmetry follows from a
general symmetry of the problem.

The velocity asymmetry $\Delta v_x$ changes if we change the
direction of one of the fields $H_z$ or $H_{ab}$. Really, in these
cases the sign of the product $n_z \partial B_z/\partial y$ in Eq.
(\ref{asym}) remains unchanged while the sign of $n_z
\partial B_y/\partial z$ changes. A simultaneous change of the
signs of the both fields $H_z$ and $H_{ab}$ does not affect the
asymmetry value $\Delta v_x$. As it follows from the experiment,
the change of the direction of one of the field changes the sign
of the velocity difference $\Delta v_x$. This means that
$|\partial B_z/\partial y| \ll |\partial B_y/\partial z|$. It is a
natural result since the pinning of the flux lines lying in $ab$
plane is usually higher than for the flux lines directed along $c$
axis. \cite{Blat,Brandt} According to Eq.~(\ref{asym}), the
velocity asymmetry $\Delta v_x$ changes its sign if the angle
between the in-plane field and the sample edge is changed from
$\alpha$ to $-\alpha$, which also is in accordance with the
experimental results.

Besides the discussed above mechanism, the Magnus
force~\cite{Blat,Brandt} could be a reason for asymmetrical creep.
However, in this case it would be impossible to explain the
observed effect of the inclination angle $\alpha$ on the creep
rate. In addition, the Magnus force should be small for an extreme
type-II superconductor. The best symmetry of
the flux creep with respect to the in-plane field direction is
attained in the experiments at inclination angles different from zero (at $\alpha$
about 4 $^\circ$). Probably this is a consequence of Magnus force influence on relaxation.

In conclusion, the MO studies of the flux creep in the Bi2212:Pb
single crystals placed in the in-plane magnetic field were
performed. At $T<54$~K, the experiments reveal a strong anisotropy
of the flux creep rate and the rate of the screening current decay
with respect to the direction of the in-plane magnetic field. This
observation confirms the existence of the strong superconducting
correlations between CuO planes in this superconductor at low
temperatures. The asymmetry of the flux creep with respect to the
in-plane field direction was also observed and explained in terms
of the interaction of the flux lines with laminar structure.

\begin{acknowledgments}

This work is supported by INTAS (grant 01--2282), RFBR (grants
02--02--17062 and 03--02--16626) and Russian State Program on
Superconductivity (projects 40.012.1.1.43.56 and
40.012.1.1.11.46).

\end{acknowledgments}

\end{document}